# Ultrahigh capacitive energy storage in highly oriented Ba(Zr$_x$Ti$_{1-x}$)O$_3$ thin films prepared by pulsed laser deposition

Alvaro A. Instan[1], Shojan P. Pavunny[1,2*], Mohan K. Bhattarai[1], and Ram S. Katiyar[1]∗

[1]Department of Physics and Institute for Functional Nanomaterials, University of Puerto Rico, P.O. Box 70377, San Juan, PR 00936-8377, USA

[2]ASEE Research Fellow at U.S. Naval Research Laboratory, 4555 Overlook Ave SW, Washington DC 20375, USA

## Abstract

We report structural, optical, temperature and frequency dependent dielectric, and energy storage properties of pulsed laser deposited (100) highly textured BaZr$_x$Ti$_{1-x}$O$_3$ (x=0.3, 0.4 and 0.5) relaxor ferroelectric thin films on La$_{0.7}$Sr$_{0.3}$MnO$_3$/MgO substrates which make this compound as a potential lead-free capacitive energy storage material for scalable electronic devices. A high dielectric constant of ~1400–3500 and a low dielectric loss of <0.025 were achieved at 10 kHz for all three compositions at ambient conditions. Ultrahigh stored and recoverable electrostatic energy densities as high as 214 ± 1 and 156 ± 1 J/cm$^3$, respectively, were demonstrated at a sustained high electric field of ~3 MV/cm with an efficiency of 72.8 ± 0.6 % in optimum 30% Zr substituted BaTiO$_3$ composition.

Future use of electricity generated from any renewable energy source will benefit greatly from advanced approaches to scalable capacitive energy storage module with a favorable figure for the power-to-mass ratio (specific power) including its usage in personal electronics, entertainment, and homeland security. The capacitive energy storage in dielectrics has been in existence for several decades allowing rapid charge storage and discharge.[1,3] Utilizing right ferroelectric material compositions with appropriate architecture, we can achieve large energy densities reducing overall weight and volume. To achieve this goal, we need to design novel energy materials and their thin film geometries that will withstand high electric fields (≥ 1 MV/cm) and maintain a high dielectric constant (>1,000s). The maximum amount of energy ($U_{ST}$) that can be stored per unit volume in ideal linear dielectric materials configured in a metal-insulator-metal (MIM) capacitor structure is given by the relation, $U_{ST} = CV^2/2 = \varepsilon_0 \varepsilon_r E_{BD}^2/2$, where, $C$ is the capacitance, $V$ is the voltage, $E_{BD}$ is the breakdown electric field strength ($E = V/d$, where $d$ is the thickness of the dielectric), $\varepsilon_0$ is the vacuum permittivity and $\varepsilon_r$ is the relative permittivity of the dielectric material being investigated. From the above equation, it is evident that the development of materials with extremely high dielectric constant and high breakdown field are the two primary requirements toward achieving high energy storage performances.

---

∗Corresponding authors: e-mail:shojanpp@gmail.com (S. P. P.), rkatiyar@hpcf.upr.edu (R. S. K). Tel.: 787 751 4210. Fax: 787 764 2571.

The maximum energy density that can be stored in a capacitor is not only a function of material composition, but it is also a function of the geometry utilized in the device. A higher energy density has largely been achieved by increasing the electric field ($E = V/d$) in the dielectric.[4,5] In thin film form the dielectric strength increases considerably, partially due to one dimension being at nanoscale which allows a better heat dissipation, keeps trapping rate below the detrapping rate and decreases the probability of finding critical defect concentrations.[6,7] The nanoscale dimension may increase Young's modulus ($Y$) as the layer becomes thinner and results in elevated breakdown electric field as this parameter follows the relation, $E_{BD} \propto Y^n$, where the exponent $n$ varies with thickness.[8] Basically, dielectric breakdown can be caused by a number of successive physical processes such as avalanche and field emission break, thermal breakdown produced by increasing conductance from Joule heating, an electromechanical collapse brought on by the attractive forces between electrodes, electrochemical deterioration, dendrite formation, etc. Out of all these a major breakdown initiator is the avalanche emission (caused by the electronic charge injection from the cathode to the conduction band of dielectric material and subsequent ionization of atoms to produce more and more electrons until conduction occurs through the bulk),[9] wherein an inverse exponential theoretical dependence on dielectric thickness is expected, $E_{BD} \propto e^{-kd}$, where, $k$ is a constant. Decreasing the thickness ($d$) of the active dielectric layer can serve to both increase capacitance and miniaturize device dimension, and is therefore a very attractive approach for low-voltage electronic applications.

Recently, energy storage capacitor and power generation research attention has turned towards lead-free and environmentally-friendly materials. Barium titanate (BaTiO$_3$) based ferroelectric materials are attractive because of their high dielectric constant, polarization, and high piezoelectric properties.[10–12] Among the doped BaTiO$_3$ systems BaZr$_x$Ti$_{1-x}$O$_3$ (BZT)[13–18] solid solutions have received renewed attention due to their high-strain level and high piezoelectric properties both in bulk and thin film forms. BaZr$_x$Ti$_{1-x}$O$_3$ exhibits a pinched phase transition at $x$~0.15, i.e., all the three phase transitions that correspond to pure BaTiO$_3$ are merged or pinched into one broad peak. With further increase in Zr concentration, a typical ferroelectric relaxor behavior has been observed.[17] Previous reports have shown the polar-cluster-like behavior in BZT solid solution having a higher concentration of Zr, $0.85 \leq x \leq 1.00$.[19,20] BZT compositions with higher concentrations of the non-ferroactive dopant ions Zr$^{4+}$ form disordered ferroelectric materials, in which polar-nanoregions (PNRs) are stabilized on cooling by random electric fields due to either charge or structural disorder.[21]

BaZr$_x$Ti$_{1-x}$O$_3$(BZT30, BZT40, and BZT50 for x=0.3,0.4, and 0.5 respectively) thin films of about 400 nm thickness were prepared by pulsed laser deposition (PLD) technique. Barium carbonate (BaCO$_3$) (99%), zirconium (IV) oxide (ZrO$_2$) (99.7%), titanium (IV) oxide (TiO$_2$) (99.8%) were used as precursors to synthesize PLD targets by solid state reaction. We used an excimer laser (KrF, 248 nm) operating at a pulse frequency of 10 Hz for BaZr$_x$Ti$_{1-x}$O$_3$/La$_{0.7}$Sr$_{0.3}$MnO$_3$ (LSMO) heterostructure thin film fabrication on MgO (100) substrates kept at 650 °C. The LSMO layer was crystallized in situ at 725 °C in 300 mTorr of oxygen for 60 min and cooled down slowly to room temperature to form highly oriented bottom electrode. During laser ablation of BZT on LSMO coated MgO at 650 °C, laser energy of 250 mJ was used and an oxygen process pressure of ~150 mTorr was preserved inside the chamber. The BZT thin films were crystallized by in situ post deposition anneal at 650 °C for 60 min in 300 mTorr of oxygen and were cooled down slowly to room temperature to form preferentially (100)-oriented BZT/LSMO heterostructures. The phase formation and crystallographic structure of the samples were confirmed with X-ray diffraction patterns, using an X-ray diffractometer (Rigaku Ultima III) operating in the Bragg-Brentano geometry outputting CuKα radiation at a wavelength of λ=1.5404 Å. The Raman measurements were performed using an ISA T64000 triple monochromator. For electrical measurements, platinum (Pt) top electrodes of ~80 μm diameter were sputter deposited through a shadow mask onto the BZT thin films to form Pt/BZT/LSMO metal-insulator-metal (MIM) capacitors. The dielectric constant of the samples was measured using an HP4294 inductance capacitance resistance (LCR) meter from 100 Hz to 1 MHz in the temperature range of 100–450 K at a 5 K/min cooling/heating rate. The remanent polarization and coercive field of the ferroelectric capacitors were measured using a Sawyer Tower test configuration (Radiant Technologies).

X-ray diffraction (XRD) patterns (Fig.S1 in the supplementary material) showed that BZT thin films have a highly (100) textured perovskite tetragonal crystal structure with reduced tetragonality, c/a, that is characterized by peak (002)/(200) splitting at 2θ ~45° as shown in the inset of Fig.S1.[22,23] It can be seen that the Bragg peak positions shift towards lower angles with an increase in % Zr content suggestive of an increase in the crystal unit cell size with the increase in Zr content incorporation as Zr is having an ionic radius of 0.98 Å greater than that of Ti, 0.72 Å. Raman-active optical modes for $BaTiO_3$ in the tetragonal phase with $P4mm$ spacegroup are 3 $A_1(TO)$ + 3 $A_1(LO)$ + 4 $E(TO)$ + 4 $E(LO)$ + 1 $B_1$.[24] The room temperature Raman spectrum of all BZT compositions displayed in Fig.S3 shows broad and mixed phonon modes at 160 $cm^{-1}$ ($E(TO_2)+E(LO_1)+A_1(TO_1)+A_1(LO_1)$), 295 $cm^{-1}$ ($E(TO_3)+E(LO_2)+B_1$), 515 $cm^{-1}$ ($E(TO_4)+A_1(TO_3)$), and 728 $cm^{-1}$ ($E(LO_4)+A_1(LO_3)$). The observed decrease in phonon mode energies and weakening of mode intensities as the Zr content increases(force constant decreases) may be considered as an indication of the incorporation of chemically more stable $Zr^{4+}$ at $Ti^{4+}$ site of the polar $BaTiO_3$ matrix and subsequent disruption of the long-range ferroelectric ordering and disordered cubic structure formation as evident from too weak Raman scattering displayed by BZT50. The dip observed at 125 $cm^{-1}$ represents B-site cation ordering and is an indication of the expected relaxor behavior.[25]

The room temperature frequency ($10^2 – 10^6$ Hz) dependence of dielectric constant and loss tangent of Pt/BZT/LSMO capacitors are shown in Fig. 1. At low frequencies (<10 kHz) changes in dielectric constant are small for BZT30, whereas for BZT40 and BZT50, dielectric constant remains almost constant. The observed fall in dielectric constant (>10 kHz) occurs due to the fact that the polarization does not arise instantaneously with the application of the electric field as mainly the relaxation of the dipole polarization commences. This type of frequency dependent dielectric behavior has been found in many ferroelectric ceramics.[26,27] We obtained a higher dielectric constant for BZT30 than for BZT40 and BZT50, this may be because of the micro-regions in BZT30 contributing to the rise in permittivity due to internal electric field distribution while in BZT40 and BZT50 these micro-regions are substantially reduced when more disorder is introduced due to the higher concentration of Zr ions. The obtained high permittivity of the BZT attributed to the dipole polarization of PNRs makes it suitable for multilayer ceramic capacitor (MLCC) applications.[28] This enhanced permittivity might be correlated to the fact that the ionic polarizability of $Zr^{4+}$ ions is higher than that of $Ti^{4+}$ ions based on the dielectric polarizability theory.[29] On the other hand, dielectric loss increases as the frequency increases. For low frequencies (<1 kHz), we can see low dielectric losses (~0.025) for all three compositions. The values of tan δ were heavily dependent on the Zr content for frequencies >1 kHz. For example, the $\varepsilon_r$ values at 10 kHz are 3446, 1971 and 1406, while the tan δ values are 0.115, 0.033 and 0.017 for BZT30, BZT40 and BZT50 thin films, respectively.

A diffuse phase transition (DPT) is present in materials where the composition fluctuations lead to large fluctuations in the Curie temperature. A DPT is generally characterized by (a) broadening in the dielectric constant (ε) versus temperature (T) curve, (b) a relatively large separation (in temperature) between the maximum of the real (dielectric constant) and imaginary (dielectric loss) parts of the dielectric spectrum, (c) a deviation from Curie–Weiss law in the vicinity of $T_m$ (transition temperature), and (d) frequency dispersion of both ε and tan δ in the transition region thereby implying a frequency dependence of $T_m$.[23,30,31] It is known that above the Curie temperature the dielectric permittivity of a normal ferroelectric follows the Curie–Weiss law formulated as $\frac{1}{\varepsilon} = \frac{(T-T_m)}{C}, T \geq T_m$ where, $T_m$ is the Curie–Weiss temperature and C is the Curie–Weiss constant. For BZT thin films, the dielectric constant (ε) was fitted to the Curie–Weiss law (solid line in Fig. 2) by plotting inverse dielectric constant (at 10 kHz) against temperature. It can be seen that a deviation from the Curie–Weiss law starting at $T_{cw}$, becomes more evident as the Zr content decreases. The Curie–Weiss law parameters determined by modeling the experimental data are listed in Table 1. The parameter $\Delta T_m$, which is used often to describe the degree of the deviation from the Curie–Weiss law, is defined as $\Delta T_m = T_{cw} - T_m$, where $T_{cw}$ describes the temperature from which the permittivity starts to deviate from the Curie–Weiss law and $T_m$ is the temperature that corresponds to the dielectric constant maxima. For BZT30, BZT40 and BZT50 the values of $\Delta T_m$ are 60, 170 and 220 K, respectively, as shown in Table 1. The values of $\Delta T_m$ increase when Zr concentration increases, which provides evidence of a composition-induced diffuse phase transition behavior in the BZT thin films with 0.30≤x≤0.50. Uchino and Nomura[32] proposed a modified empirical expression to describe the diffuseness of ferroelectric phase transition as $\frac{1}{\varepsilon} - \frac{1}{\varepsilon_m} = \frac{(T-T_m)^\gamma}{C'}$ where γ is the degree of relaxation and C' is assumed to be constant. The parameter γ gives information on the character of the phase transition: for γ = 1, a normal Curie–Weiss law is obtained which is valid for a perfect ferroelectric, γ = 2 describes a complete diffuse phase transition for a perfect relaxor material,[23,33] γ value between 1 and 2 corresponds to the incomplete diffuse phase transition where the correlated ferroelectric clusters are hypothesized. The inset of Fig. 2 shows the linear relationship between ln (1/ε-1/$\varepsilon_m$) as a function of ln (T-$T_m$) for the three samples. The slope of the linear fit of the curves is used to determine the γ values as 1.83, 2.02 and 1.85 for BZT30, BZT40 and BZT50 thin films, respectively, and shows a strong diffuse phase transition or relaxor behavior.

Figure 3 shows the non-saturated slim room temperature hysteresis loops for all BZT thin films measured in Pt/BZT/LSMO capacitor geometry that sustained high electric fields of ~3 MV/cm at a frequency of 10 kHz. The observed behavior in this slimmer P-E loop possibly indicates that the size of the polar regions is smaller than the dimension of the micron sized domains in the thin films, but, it is large enough to have significant dipolar cooperation among neighboring unit cells to present polarization hysteresis.[34] As zirconium (*Zr*) and titanium (*Ti*) have the same valency and only differ in their ionic radius ($R_{Zr}^{4+}$ = 0.98 Å, $R_{Ti}^{4+}$ = 0.72 Å), it is expected that the addition of small amounts of zirconium in barium titanate does not generate sufficient disorder. However, when adequate quantities of zirconium (0.30 ≤x ≤0.50 in this study) are added, an interaction of polar and non-polar regions is generated resulting in the relaxor behavior with low coercive fields (~0.28 MV/cm) indicating that the thin films are soft electrically. The remanent polarization ($P_r$) decreases gradually from 38 µC/cm$^2$ to 32 µC/cm$^2$ and then to 29 µC/cm$^2$ with the increase in *Zr* content (See Table 2). This reduction in the remanent polarization of BZT thin films can be explained by the difference of the radius of $Zr^{4+}$ and $Ti^{4+}$.

The stored electrostatic energy densities, ($U_{ST} = \int_0^{P_{MAX}} EdP$ where, $P_{MAX}$ is the charge density or the polarization measured at the maximum electric field $E_{MAX}$ represented by region I + II in the inset of Fig. 3) recoverable energy densities ($U_{RE} = \int_{P_r}^{P_{MAX}} EdP$ region I in the inset of Fig. 3) and charge-discharge efficiency ($\eta = (U_{RE}/U_{ST}) \times 100$) determined for Pt/BZT/LSMO capacitors are given in Table 2. It was found that energy density increased with increase in electric field and decrease in Zr content for all BaZr$_x$Ti$_{1-x}$O$_3$ (0.30≤x≤0.50) thin films. In particular, $U_{RE}$ was found to increase with decreasing Zr content from 144 ± 0.7 J/cm$^3$ at x=0.50 to 156 ± 0.7 J/cm$^3$ at x=0.30. These ultrahigh energy density values suggest the extremely high potential of BZT thin films to store electrical energy. The charge-discharge efficiency of these thin films varied from 72.8 ± 0.6 % to 82.4 ± 1.0 % at an electric field around 3.0 MV/cm and these figures are very high compared to reports on different compositions.[35–37] In our current work, the successive evolution of relaxation in the BaZr$_x$Ti$_{1-x}$O$_3$ (BZT) thin films has been hypothesized as being due to the increasing amount of ordering and density of nano size Ti$^{4+}$ rich polar regions in the Zr$^{4+}$ rich matrix as Ti$^{4+}$ is progressively incorporated in the BaZrO$_3$ lattice. It is also seen that remanent polarization decreases with increase in Zr content. This could be due to the fact that Zr content can form BaZrO$_3$ cubic structure and therefore less polar regions can be found in the film getting low remanent polarization.

We have fabricated(100) highly textured BaZr$_x$Ti$_{1-x}$O$_3$ (0.30≤x≤0.50) thin films on LSMO/MgO substrates utilizing optimized PLD process and demonstrated them as potential lead-free capacitive energy storage materials for scalable electronic devices in terms of their structural, optical, temperature and frequency dependent dielectric, and ferroelectric charge storage properties. Ultrahigh stored and recoverable electrostatic energy densities as high as 214 ± 1 and 156 ± 1 J/cm$^3$, respectively, were achieved at a sustained high electric field of ~3 MV/cm with an efficiency of 72.8 ± 0.6 % in optimum 30% Zr substituted BaTiO$_3$ composition.

See supplementary material for XRD, AFM and Raman data of the studied BZT samples.

**Acknowledgements**

Financial support from DOD Grant No. FA9550-16-1-0295is acknowledged. SPP gratefully acknowledges the financial assistance from IFN-NSF under the Grant No. EPS-1002410.

**Figures:**

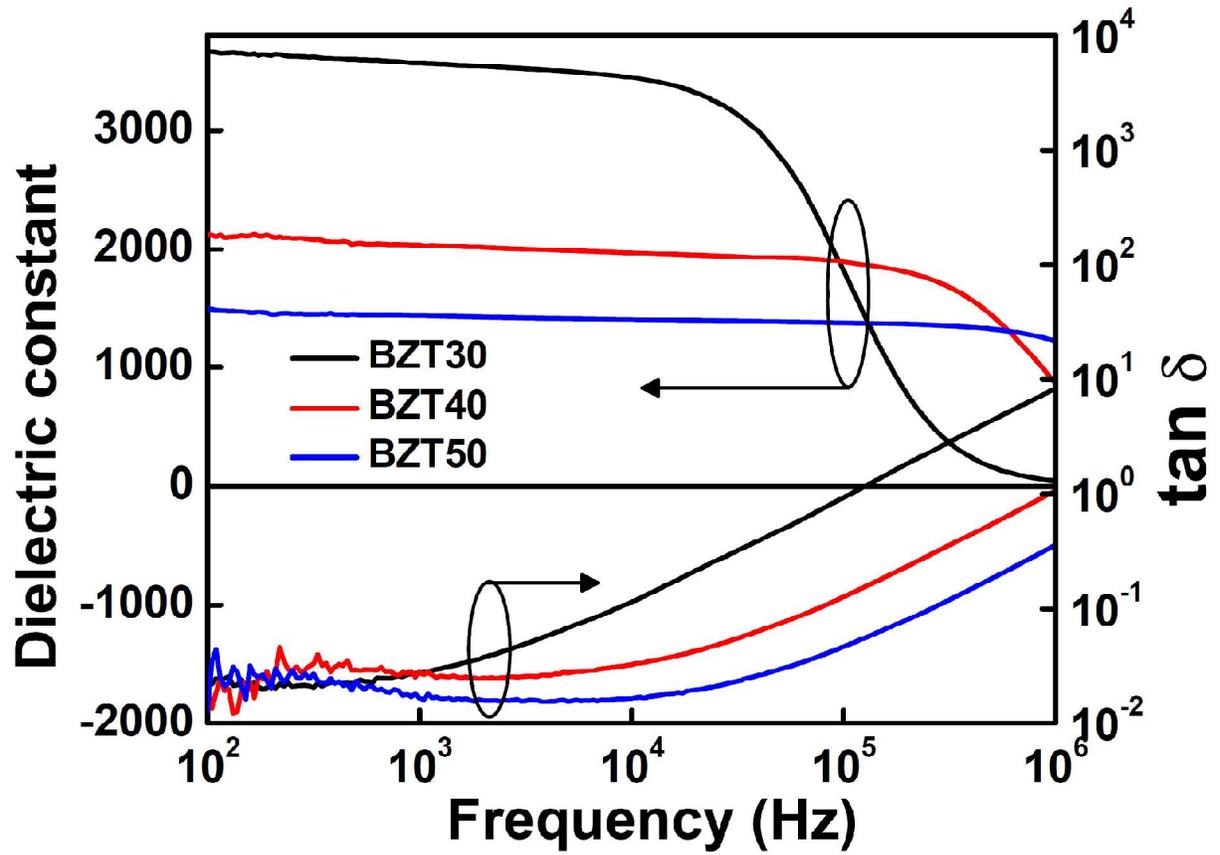

Fig.1

Fig. 1. Frequency dependence of dielectric constant and loss tangent of BZT thin films at room temperature.

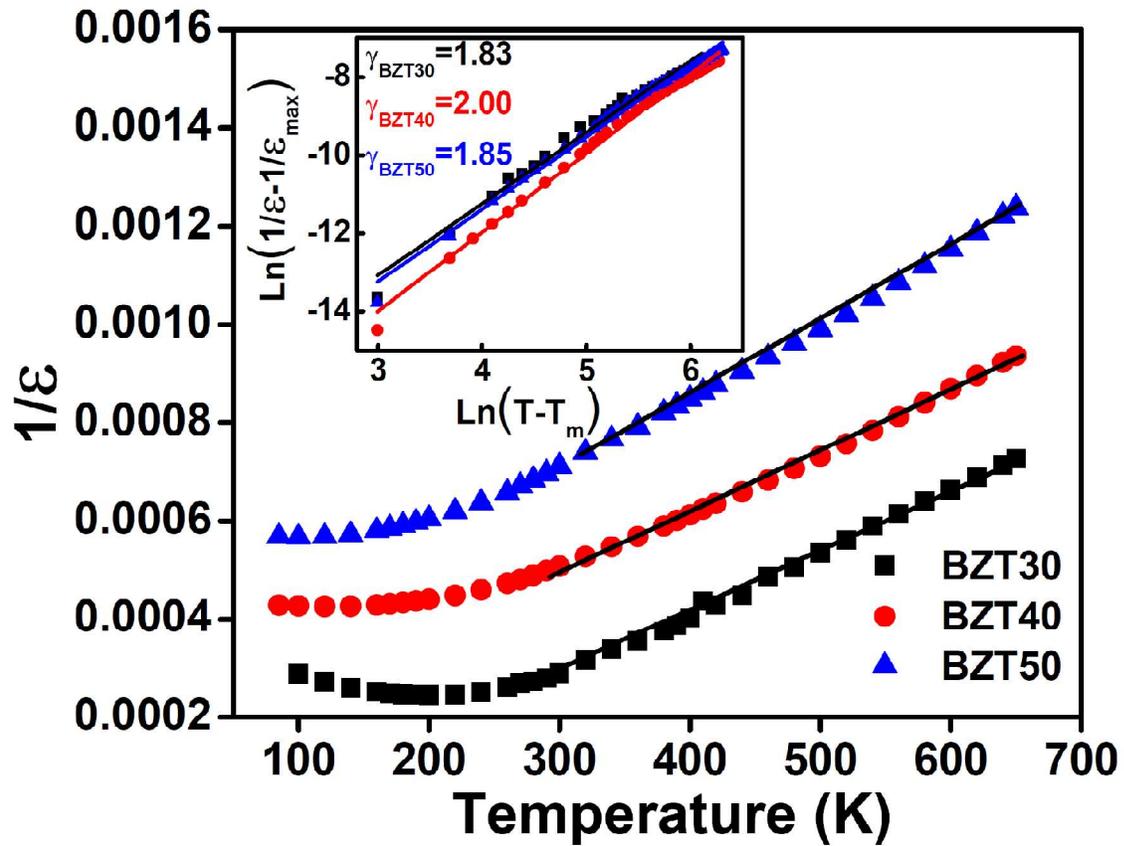

Fig.2

Fig. 2. The inverse dielectric constant (1/ε) as a function of temperature at 10 kHz for the three samples. The inset shows their Ln (1/ε−1/ε$_m$) vs ln (T−T$_m$) plots. (Symbols: experimental data, solid line: Curie-Wiss fits).

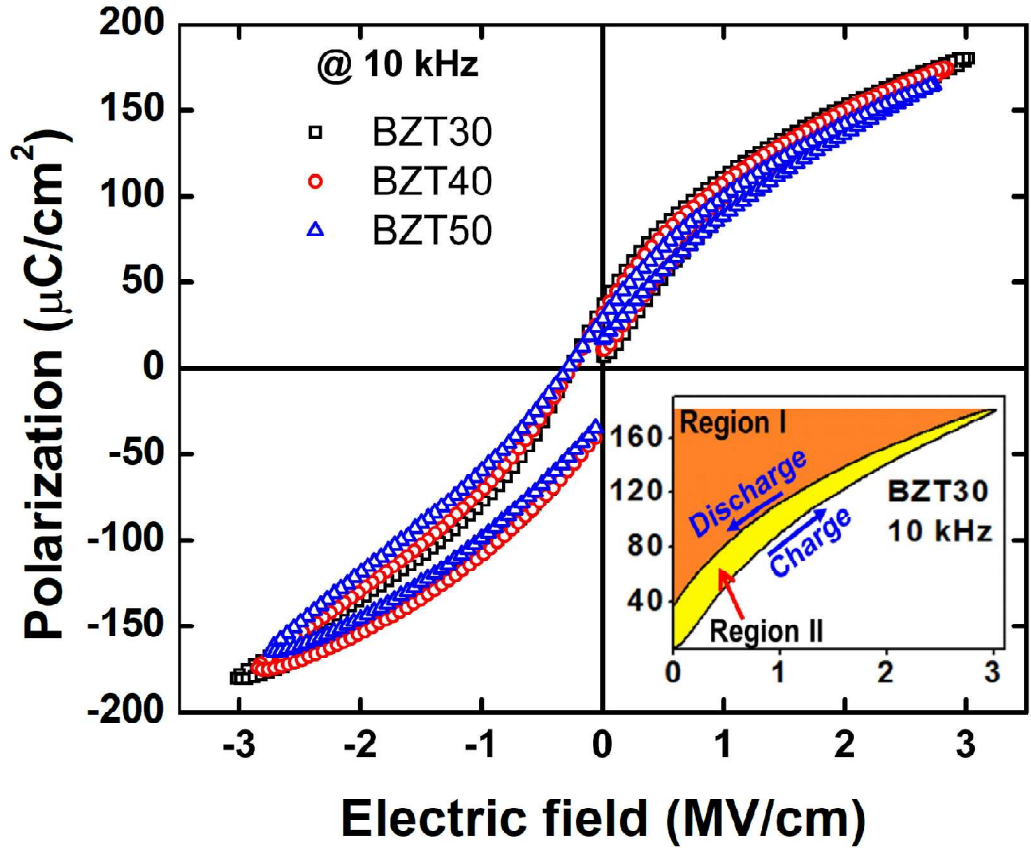

Fig. 3. Room-temperature ferroelectric hysteresis loops of $BaZr_xTi_{1-x}O_3$ (0.30≤x≤0.50) (BZT) thin films as a function of electric field. The inset shows the charge-discharge cycle of the $BaZr_{0.3}Ti_{0.7}O_3$ thin film at 10 kHz and the related electrical energy densities: area I (orange shaded area) corresponds to the discharged energy density and area II (yellow shaded area) is the energy loss density.

Table 1: The maximum dielectric constant ($\varepsilon_m$), the Curie–Weiss temperature ($T_m$), the Curie–Weiss-constant (C), and the temperature above which the dielectric constant follows the Curie–Weiss law ($T_{cw}$), $T_{cw}-T_m$, and the critical parameter $\gamma$ for the three BZT samples at 10 kHz.

| Sample | $\varepsilon_m$ | $T_m$ (K) | C (x$10^5$K) | $T_{cw}$(K) | $T_{cw}-T_m$ (K) | $\gamma$ |
|---|---|---|---|---|---|---|
| **BZT30** | 4077 | 200 | 8.1 | 260 | 60 | 1.83 |
| **BZT40** | 2346 | 120 | 7.9 | 290 | 170 | 2.02 |
| **BZT50** | 1765 | 100 | 6.6 | 320 | 220 | 1.85 |

Table 2: Energy density values for the three BZT thin film compositions.

| Thin Film | $E_c$ (MV/cm) | $P_r$ ($\mu C/cm^2$) | $U_{ST}$ (J/cm$^3$) | $U_{RE}$ (J/cm$^3$) | Efficiency ($U_{RE}/U_{ST}$)*100% |
|---|---|---|---|---|---|
| BZT30 | 0.28 | 39 | 214 ± 1 | 156 ± 0.7 | 72.8 ± 0.6 |
| BZT40 | 0.26 | 33 | 192 ± 0.5 | 150 ± 1.0 | 78.3 ± 0.5 |
| BZT50 | 0.31 | 31 | 175 ± 1.1 | 144 ± 0.7 | 82.4 ± 1.0 |

# Supplementary Materials

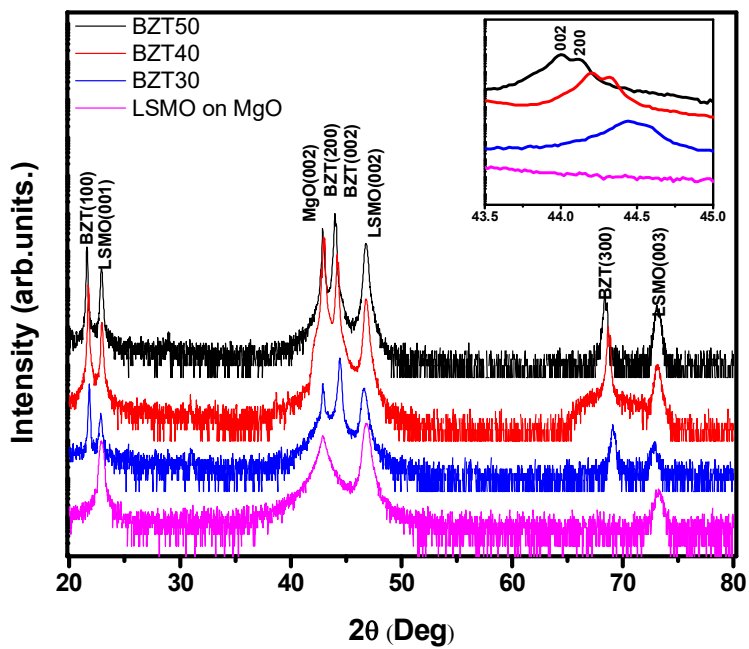

**Fig. S1.** Room temperature XRD patterns of LSMO coated on MgO (100) substrate and BZT thin films with various Zr content on LSMO coated MgO (100) substrates.

The surface morphology and roughness were determined by a NanoScope IIIA (Digital Instruments, USA) atomic force microscope (AFM) using tapping mode amplitude modulation. Representative AFM micrographs of BZT thin films recorded in contact mode with a scan size of 3x3 µm$^2$ are inserted in Fig. S2. It is evident that these films have developed well dense granular morphology with an average grain size of about 15 nm and possess a smooth and homogeneous surface topography with an average surface roughness of about 3 nm. The $R_{rms}$ values obtained from AFM image for BZT/LSMO samples were small for all the compositions (e.g. $R_{rms}$(BZT30=3.80 nm), $R_{rms}$ (BZT40=0.45 nm), $R_{rms}$ (BZT50=2.02 nm)) and make them suitable for applications in energy storage.

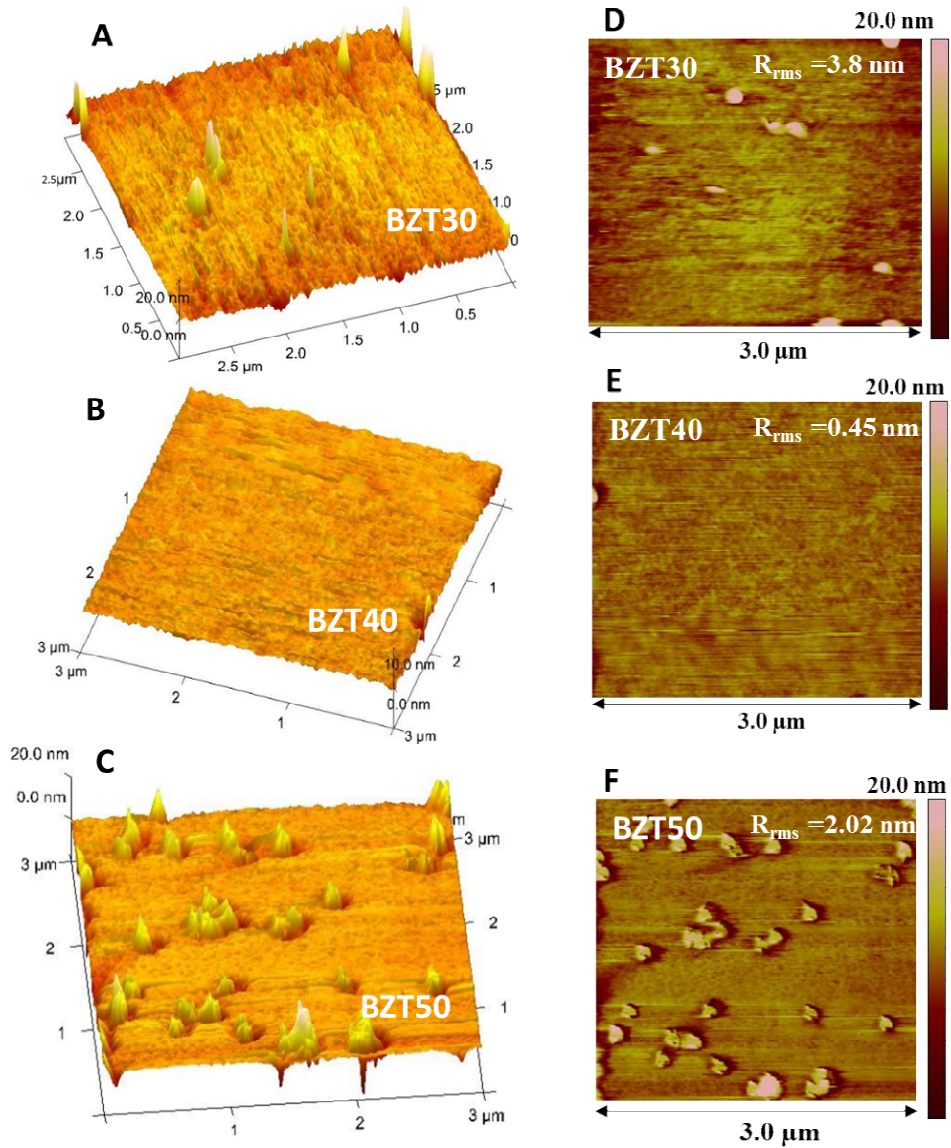

**Fig. S2.** Atomic force microscopy (AFM) surface topography images, of 3 × 3 $\mu m^2$ having 20 nm in Z-scale of $BaZr_xTi_{1-x}O_3$ (0.30≤x≤0.50) (BZT) thin films: (A) and (D) x=0.30, (B) and (E) x=0.40, (C) and (F) x=0.50 deposited on LSMO/MgO.

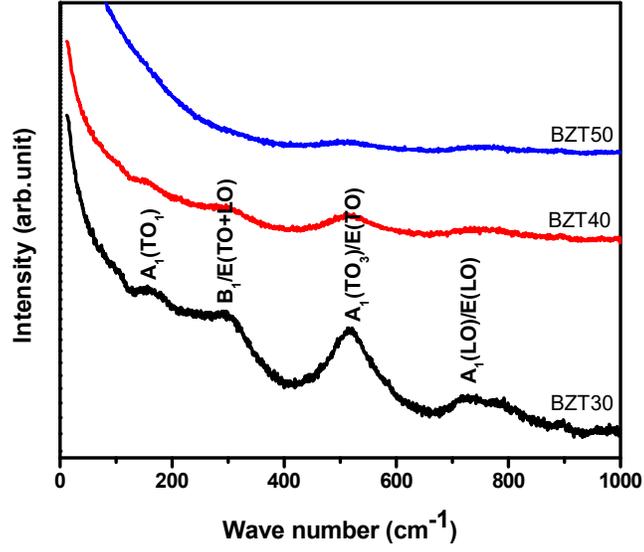

**Fig. S3.** Room-temperature Raman spectra of BaZr$_x$Ti$_{1-x}$O$_3$ (0.30≤x≤0.50) (BZT) thin films.